\documentclass{article}[9pt]
\usepackage{spconf,amsmath,epsfig,flushend,subfig,color}
\usepackage{amssymb}
\ninept 

\title{Adaptive Iterative Decision Feedback Detection Algorithms for Multi-User MIMO Systems \vspace{-0.75em}}
\name{ Peng~Li,~Jingjing~ Liu and Rodrigo~ C.~ de~ Lamare
\vspace{-1em}}
\address{\small Communications Research Group, Department of Electronics,
University of York, UK \\
\small Email: \{pl534,jl622,rcdl500\}@ohm.york.ac.uk
\vspace{-1.25em}}
\begin{document}

\maketitle
\begin{abstract}
An adaptive iterative decision multi-feedback detection algorithm
with constellation constraints is proposed for multiuser
multi-antenna systems. An enhanced detection and interference
cancellation is performed by introducing multiple constellation
points as decision candidates. A complexity reduction strategy is
developed to avoid redundant processing with reliable decisions
along with an adaptive recursive least squares algorithm for
time-varying channels. An iterative detection and decoding scheme is
also considered with the proposed detection algorithm. Simulations
show that the proposed technique has a complexity as low as the
conventional decision feedback detector while it obtains a
performance close to the maximum likelihood detector.
\vspace{-0.25em}
\end{abstract}
\begin{keywords}
MIMO systems, decision feedback receivers, RLS algorithms,
multi-user detection, iterative processing. \vspace{-0.5em}
\end{keywords}
\section{Introduction}
\label{sec:intro}

Multi-user detection (MUD) algorithms have shown that they can be
applied to $3$G and next generation multi-antenna communication
systems \cite {verdu}. As the optimal maximum likelihood detector
(MLD) has an exponential computational cost in the number of users
and constellation points, cost-effective solutions such as the
sphere decoder (SD) and decision feedback (DF) receivers
\cite{ginis,Foschini3} are preferred as they offer an acceptable
performance and complexity trade-off in spatial multiplexing
multi-input multi-output (MIMO) systems. For time-varying channels,
adaptive DF structures \cite{choi,peng,delamaretvt} are promising as
adaptive algorithms can be used to track the channels and to avoid
excessive computations when the channels are time-varying. However,
the performance of DF techniques are far from the MLD.

In this paper, an adaptive decision feedback based algorithm is
proposed for signal detection in multi-user MIMO (MU-MIMO) systems
with time-varying channels. The proposed DF algorithm can reduce the
performance gap between the optimal MLD and existing DF algorithms.
The proposed DF algorithm exploits multiple constellation points and
orderings to obtain several detection candidates. A reliability
checking technique called constellation constraint (CC) brings
improved performance to the proposed DF detector at a small
additional computational cost as compared to the conventional DF. We
also consider an iterative detection and decoding (IDD) scheme in
which the proposed DF detector is incorporated.

This paper is organized as follows: Section $2$ gives the data and
system model of the MU-MIMO system; the proposed detection scheme is
described in Section $3$, whereas the IDD scheme is detailed in
Section $4$; the simulation results are shown in Section $5$ and
Section $6$ presents the conclusions of the paper.

\section{Data and System Model}
\label{sec:data}

Let us consider a model of an uplink MU-MIMO system with $K$ users.
Each user is equipped with a single antenna. At the receiver side,
$N_R$ receive antennas are available for collecting the signals.
Throughout this paper, the complex baseband notation is used while
vectors and matrices are written in lower-case and upper-case
boldface, respectively. At each time instant $[i]$, $K$ users
simultaneously transmit $K$ symbols organized into a vector
${\boldsymbol s}[i] = \big[ s_1[i], ~s_2[i], ~ \ldots,~ s_{K}[i]
\big]^T$, where $(\cdot)^T$ denotes the transpose operation, and
whose entries are chosen from a complex $C$-ary constellation set
$\mathcal{A} = \{ a_1,~a_2,~\ldots,~a_C \}$. The symbol vector
${\boldsymbol s}[i]$ is transmitted over time-varying channels and
the received signal is processed by $N_R$ antennas. The received
signal is collected to form an $N_R \times 1$ vector with sufficient
statistics for detection \vspace{-0.35em}
\begin{equation}
{\boldsymbol r}[i] = \sum_{k=1}^K{\boldsymbol h}_k[i]s_k[i] +
{\boldsymbol v}[i] = {\boldsymbol H}[i] {\boldsymbol s}[i] +
{\boldsymbol v}[i],
\end{equation}\vspace{-0.15em}where the $N_R \times 1$ vector ${\boldsymbol v}[i]$ represents a
zero mean complex circular symmetric Gaussian noise with covariance
matrix $E\big[ {\boldsymbol v}[i] {\boldsymbol v}^H[i] \big] =
\sigma_v^2 {\boldsymbol I}$, $\sigma_v^2$ is the noise variance and
${\boldsymbol I}$ is the identity matrix, $E[ \cdot]$ stands for the
expected value and $(\cdot)^H$ denotes the Hermitian operator. The
symbol vector ${\boldsymbol s}[i]$ has zero mean and a covariance
matrix $E\big[ {\boldsymbol s}[i] {\boldsymbol s}^H[i]\big] =
\sigma_s^2 {\boldsymbol I}$, where $\sigma_s^2$ is the signal power.
Furthermore, the elements of $\boldsymbol{H}[i]$ are the
time-varying complex channel gains from the $n_T$-th transmit
antenna to the $n_R$-th receive antenna, which follow the Jakes'
model \cite{Jakes}. The $N_R \times 1$ vector $\boldsymbol h_k[i]$
includes the channel coefficients of user $k$ such that
${\boldsymbol H}[i]$ is formed by the channel vectors of all users.
As the optimal SINR-based nulling and cancellation order (NCO)
\cite{choi} requires a high computational complexity, we determine
the NCO by computing the norms of the column vectors corresponding
to all users and we then detect them in decreasing order of their
norms. 

\section{Proposed Adaptive Multi-user DF Detector}
\label{sec:proposed}
In the proposed adaptive multi-user DF detector, called AMUDFCC, the
received signal $\boldsymbol{r}[i]$ is filtered by a $N_R \times 1$
forward filter $\boldsymbol{\omega}^H_{f,k}[i]$ which acts as the
nulling vectors of the V-BLAST algorithm. Then for each user stream
$k = 1, \ldots, K$, the decisions are accumulated and cancelled by
the $(k-1)$-dimensional decision backward filter
$\boldsymbol{\omega}^H_{b,k}[i]$. Let $\hat{\boldsymbol{s}}[i] =
\big{[}\hat{s}_1[i],\hat{s}_2[i],\ldots,\hat{s}_{K}[i]\big{]}^T$
represent the detected symbol vector and $u_k[i]$ denotes the
difference between the forward filter output and the backward filter
output as described as
\begin{equation}\label{df}
u_k[i] = \boldsymbol{\omega}^H_{f,k}[i]\boldsymbol{r}[i] -
\boldsymbol{\omega}^H_{b,k}[i]\hat{\boldsymbol{s}}_{k-1}[i],
\end{equation}
where $\boldsymbol{\omega}_{b,1}^H = \boldsymbol{0} $ for the first
user and the $(k-1)$-dimensional detected symbol vector is defined
as
\begin{equation}
\hat{\boldsymbol{s}}_{k-1}[i]=\big{[}\hat{s}_1,\hat{s}_2,\ldots,\hat{s}_{k-1}\big{]}^T.
\end{equation}
For notational convenience, the feedforward and feedback filters can
be concatenated together as \cite{choi}
\begin{equation}
\tilde{\boldsymbol{\omega}}_k[i] =
\begin{cases}
\boldsymbol{\omega}_{f,k}[i], &  k=1\\
\big{[}\boldsymbol{\omega}_{f,k}^T[i],\boldsymbol{\omega}_{b,k}^T[i]\big{]}^T,
&  k=2,\ldots, K.
\end{cases}
\end{equation}
The input can also be concatenated as
\begin{equation}\label{concat2}
\tilde{\boldsymbol{r}}_k[i] =
\begin{cases}
\boldsymbol{r}[i], &  k=1\\
\big{[}\boldsymbol{r}^T[i]-\hat{\boldsymbol{s}}_{k-1}^T[i]\big{]}^T,
&  k=2,\ldots, K.
\end{cases}
\end{equation}
Then, we can rewrite (\ref{df}) as
\begin{equation}\label{concatenated}
u_{k}[i] =
\tilde{\boldsymbol{\omega}}_k^H[i]\tilde{\boldsymbol{r}}_k[i].
\end{equation}

\begin{figure}[!htb]
\begin{center}
\def\epsfsize#1#2{0.85\columnwidth}
\epsfbox{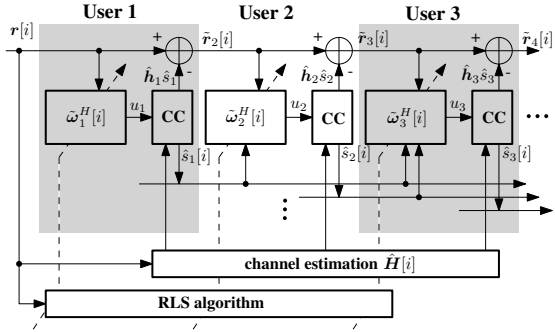} \vspace{-1em} \caption{\footnotesize Block
diagram of the proposed AMUDFCC detector. }
\vspace{-1em}\label{block}
\end{center}
\end{figure}
As a result, the structure and the signal processing model of the
proposed DF detector are depicted in Fig.{\ref{block}}. We denote
the receive filter of each user as
$\tilde{\boldsymbol{\omega}}_k^H[i]$ ($k = 1, \ldots, K$), and the
value of each entry can be obtained by solving the standard least
squares (LS) problem. The LS cost function with an exponential
window is given by
\begin{equation}\label{lsp}
\mathcal{J}_k[i] = \sum_{\tau=1}^{i} \lambda^{i-\tau}
\Big{|}\hat{s}_k[\tau]-\tilde{\boldsymbol{\omega}}^H_k[i]\tilde{\boldsymbol{r}}_k[\tau]\Big{|}^2,
\end{equation}
where $0\ll\lambda<1$ is the forgetting factor, the scalar $\hat{s}_k[\tau]$ denotes the detected
signal in the time index $\tau$ or the known pilots where $\hat{s}_k[\tau] = s_k[\tau]$. The optimal tap
weight minimizing $\mathcal{J}_k[i]$ is given by
\begin{equation}\label{ls}
\tilde{\boldsymbol{\omega}}_k[i] =
\boldsymbol{\Phi}_k^{-1}[i]\boldsymbol{p}_k[i],
\end{equation}
where the time-averaged cross correlation matrix is obtained by
$\boldsymbol{\Phi}_k[i] = \sum_{\tau=1}^{i}
\lambda^{i-\tau}\tilde{\boldsymbol{r}}_k[\tau]\tilde{\boldsymbol{r}}_k^H[\tau]$
and $\boldsymbol{\Phi}_k[0] = \boldsymbol{0}$,
the time-averaged cross correlation vector is defined by
$\boldsymbol{p}_k[i] = \sum_{\tau=1}^{i}
\lambda^{i-\tau}\tilde{\boldsymbol{r}}_k[\tau]\hat{{s}}_k^*[\tau].$

Using the recursive least squares (RLS) algorithm \cite{haykin}, the
optimal weights in (\ref{ls}) can be calculated recursively as
follows:
\begin{equation}\label{rls1}
\boldsymbol{q}_k[i]=\boldsymbol{\Phi}_k^{-1}[i-1]\boldsymbol{r}_{k}[i],
\end{equation}
\begin{equation}\label{rls2}
\boldsymbol{k}_k[i]=\frac{\lambda^{-1}\boldsymbol{q}_k[i]}{1+\lambda^{-1}\boldsymbol{r}_{k}^H[i]\boldsymbol{q}_{k}[i]},
\end{equation}
\begin{equation}\label{rls3}
\boldsymbol{\Phi}_k^{-1}[i]=\lambda^{-1}\boldsymbol{\Phi}_k^{-1}[i-1]-\lambda^{-1}\boldsymbol{k}_k[i]\boldsymbol{q}^H_k[i],
\end{equation}
\begin{equation}\label{rls4}
\tilde{\boldsymbol{\omega}}_{k}[i]=\tilde{\boldsymbol{\omega}}_{k}[i-1]+\boldsymbol{k}_k[i]\xi_k^*[i],\\
\end{equation}
where
\begin{equation}\label{rls5}
\xi_k[i]=
\begin{cases}
{s}_{k}[i] - \tilde{\boldsymbol{\omega}}_{k}^H[i-1]\tilde{\boldsymbol{r}}_{k}[i],   & \mbox{Training Mode,} \\
\hat {s}_{k}[i] - \tilde{\boldsymbol{\omega}}_{k}^H[i-1]\tilde{\boldsymbol{r}}_{k}[i],    & \mbox{Decision-directed Mode.}  \\
\end{cases}
\end{equation}
As indicated in (\ref{rls5}), this adaptive detection algorithm
works in two modes. The first one is employed with the training
sequence, while the second one is the decision-directed mode that is
switched on after the filter weights converge. In the
decision-directed mode the quality of the detected symbols has a
major impact on the performance of adaptive DF algorithms. This is
because the detection error of the current user may propagate
throughout the detection of the following users. Moreover, in
time-varying channels a poor $\xi_k[i]$ can easily damage the
$\tilde{\boldsymbol{\omega}}_k[i]$ in equation ({\ref{rls4}})
resulting in burst errors.

\subsection{Constellation Constraints}

When the filter output $u_{k}[i]$ is considered unreliable, the CC
scheme produces a number of selected constellation points as the
candidate decisions. A selection algorithm is introduced to prevent
the search space from growing exponentially, saving computational
complexity by avoiding redundant processing with reliable decisions.
\begin{figure}[!htb]
\begin{center}
\def\epsfsize#1#2{0.5\columnwidth}
\epsfbox{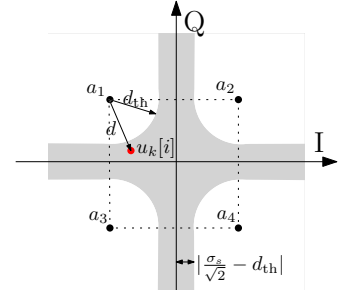} \vspace{-2.5em} \caption{\footnotesize The
constellation constraints (CC) device. The CC procedure is invoked
as the soft estimates $u_k[i]$ drop into the shaded area.}\label{cc}
\vspace{-1em}
\end{center}
\end{figure}
In the decision-directed mode, the concatenated filter output
$u_k[i]$ is checked by the CC device which is illustrated in
Fig.{\ref{cc}}, where a threshold $d_{\scriptsize \mbox{th}}$ is
defined which can be either a constant or a linear function of
$\sigma_v$. The CC device finds the nearest constellation point to
$u_k[i]$ according to
\begin{equation}
a_k[i]=\arg \min_{a_c\in\mathcal{A}}\big\{|u_{k}[i]-a_c|\big\},
\end{equation}
where $a_c$ represents all potential constellation points. A
decision is considered unreliable if at least one of the following
conditions holds
\begin{equation}\label{sa1}
d > d_{\scriptsize \mbox{th}} \qquad \mbox{when} \qquad
\begin{cases}
\big{|}\text{Re}\{u_k[i]\}\big{|} \leq \frac{\sigma_s}{\sqrt{2}} \\
\big{|}\text{Im}\{u_k[i]\}\big{|} \leq \frac{\sigma_s}{\sqrt{2}} \\
\end{cases}
\end{equation}
\begin{equation}\label{sa2}
\begin{array}{l}
\big{|}\text{Re}\{u_k[i]\}\big{|} < \frac{\sigma_s}{\sqrt{2}} - d_{\scriptsize \mbox{th}}\\
\qquad \mbox{OR}\\
\big{|}\text{Im}\{u_k[i]\}\big{|} < \frac{\sigma_s}{\sqrt{2}} - d_{\scriptsize \mbox{th}}\\
\end{array}
\mbox{when} \left\{\begin{array}{l}
\big{|}\text{Re}\{u_k[i]\}\big{|} > \frac{\sigma_s}{\sqrt{2}} \\
\big{|}\text{Im}\{u_k[i]\}\big{|} > \frac{\sigma_s}{\sqrt{2}} \\
\end{array}\right.
\end{equation}
where $d$ denotes the distance between the estimated symbol $u_k[i]$
and its nearest constellation point $a_k[i]$ \footnote{Equation
(\ref{sa2}) defines the shadowed area inside the square obtained by
connecting the four $a_c$ ($a_c = (\pm{\sigma_s} / \sqrt{2}, \pm
j{\sigma_s} / \sqrt{2})$). Equation (\ref{sa1}) denotes the shadowed
area outside the square. This concept can be further extended to
multi-tier constellations, eg. 16-QAM. }. Instead of finding the
closest vector, in fact, the scalar constellation helps to reduce
the cost. Since the CC device distinguishes whether the feedback
signal is reliable, the detector maintains its complexity at the
same level of the conventional DF structure. Once the filter output
$u_k[i]$ drops into the lighted area of the constellation map, the
decision is considered reliable and the quantization operation
$\mbox{Q}(\cdot)$ is then performed
\begin{equation}
\hat{s}_k[i] = \mbox{Q}(u_k[i]).
\end{equation}
If $u_k[i]$ drops into the shadowed area, the decision is determined
unreliable. The CC processing is evoked and a candidate vector is
generated as $\mathcal{L}=\{c_{1}, c_{2},\ldots,c_m,\ldots,c_{M}\}
\subseteq \mathcal{A}$. The candidates are constrained by the
constellation map and the selected vector is a selection of the $M$
nearest constellation points to the $u_k[i]$. The size of
$\mathcal{L}$ can be either fixed or variable, which introduces a
trade off between the performance and complexity.

The refined estimate is obtained by $\hat{s}_k[i] = c_{\scriptsize
\mbox{opt}}$ where $c_{\scriptsize \mbox{opt}}$ is the optimal
candidate selected from $\mathcal{L}$. This refined decision will
produce a more accurate $\xi_k[i]$ which minimizes the mean square
error (MSE). The benefits offered by the CC algorithm are based on
the assumption that the optimal feedback candidate $c_{\scriptsize
\mbox{opt}}$ is correctly selected. This selection algorithm is
described as follows: a set of tentative decision vectors
$\boldsymbol{B}_k = {\big{\{}}\boldsymbol{b}^1_k, \ldots,
\boldsymbol{b}^m_k, \ldots, \boldsymbol{b}^M_k{\big{\}}}$ is defined
and the number of tentative decision vectors $M$ equal the number of
selected constellation candidates. Each vector $\boldsymbol{b}^m_k$
is defined as $\boldsymbol{b}^m_k[i] =
\big{[}\hat{s}_1[i],\ldots,\hat{s}_{k-1}[i],c_m,\hat{b}_{k+1}[i],\ldots,\hat{b}_{K}[i]\big{]}$,
the $K \times 1$ vector $\boldsymbol{b}^{m}_k $ consists of: 1)
$(k-1)$-dimensional detected symbol vector
$\hat{\boldsymbol{s}}_{k-1}[i]$ which is used in (\ref{concat2}); 2)
a candidate symbol $c_m$ taken from $\mathcal{L}$ for substituting
the unreliable ${{\mbox{Q}}}(u_k[i])$ of the $k$-th data stream; 3)
by combining 1) and 2) as the previous decisions, the tentative
decisions of the following streams
$\hat{b}_{k+1}[i],\ldots,\hat{b}_{K}[i]$ are subsequently obtained
by the adaptive detector. Let us define the vector with the
candidate constellation point as
\begin{align}
\check{\boldsymbol{s}}_{k,m}[i] & = \big{[}\hat{s}_1[i],\ldots,\hat{s}_{k-1}[i],c_m\big{]}^T,\\
                & = \big{[}\hat{\boldsymbol{s}}_{k-1}^T[i],c_m\big{]}^T.
\end{align}
Therefore, (\ref{concat2}) turns out to be
\begin{equation}
\bar{\boldsymbol{r}}_{k+1,m}[i] =
\big{[}\boldsymbol{r}^T[i],\check{\boldsymbol{s}}_{k,m}^T[i]\big{]}^T,
k=1,\ldots,K.
\end{equation}
The tentative decision of the $(k+1)$ stream becomes
\begin{equation}
\hat{b}_{k+1}[i] = {\mbox{Q}\Big{\{}}
\tilde{\boldsymbol{\omega}}_{k+1}^H[i]\bar{\boldsymbol{r}}_{k+1,m}[i]\Big{\}}.
\end{equation}
The CC algorithm selects the best constellation point among $M$
candidates according to the maximum likelihood (ML) rule as
\begin{equation}\label{ED}
m_{\scriptsize \mbox{opt}}=\arg \min_{1 \leq m \leq M} \Big{\|}
\boldsymbol{r}[i]-{\hat{\boldsymbol{H}}}\boldsymbol{b}_k^m[i]\Big{\|}^2.
\end{equation}
Then $c_{\scriptsize \mbox{opt}}$ replaces the unreliable decision
$u_k[i]$. The same receive filter $\boldsymbol{\omega}_k[i]$ is used
to process all the candidates, which allows the proposed algorithm
to have the simplicity of the adaptive DF detector. Here we employ
an RLS algorithm to estimate the channel \cite {E07}.

%

\subsection{Computational Complexity}

Let us define the parameter $K=N_R$, and $M$ as the number of
candidates. The numbers of complex multiplications, corresponding to
the V-BLAST and the DF-RLS, are $2K^3+K^2+K$ and
$\frac{28}{3}K^2-\frac{4}{3}$. respectively. As for the proposed
scheme, in the worst case\footnote{We have the worst case and the
best case which means all $K$ decisions are considered unreliable
and all decisions are reliable, respectively. }, it requires
$M(5/2K^2-3/2K)$ multiplications on top of the DF algorithm. The
additional complexity is obtained by:
\begin{itemize}
\item  If $u_1$ is unreliable, we replace \mbox{Q}$(u_1)$ with
$c_m$, the multiplication repeats $M$ times for the different $c_m$.
The number of the complex multiplication is $M \times
\sum_{k=1}^{K-1} k$.

\item If $u_2$ is unreliable, as previously, the number of
complex multiplications is $1 + M \times \sum_{k=1}^{K-2} k $.

\item If $u_3$ is unreliable, the number of complex multiplications
is $2 + M \times \sum_{k=1}^{K-3} k $.

\item By summing across $K$ users we have: $\sum_{k=1}^{K}(k-1)+M\sum_{k=1}^{K-k}k.$
\end{itemize}
The overall additional complexity can be obtained by summing the
above figures with the complexity required by the ML selection rule
and the reliability checking algorithm. Moreover, the probability of
unreliable estimates decreases as the number of users
increases\footnote{This is due to the increased overall detection
diversity.}, which leads to the processing of $6.1\%$, $4.65\%$,
$3.59\%$ on average over the users of the estimated symbol for $K =
2, 4, 8$ users, respectively. The numerical results suggest that
extra computations can be further reduced in larger systems where
both $N_R$ and $K$ are larger.

\subsection{Multiple-Branch Processing}

In this subsection, the proposed detector is applied with several
parallel branches that are equipped with different NCO patterns. Let
us define $\hat{\boldsymbol{s}}'[i] \triangleq
\boldsymbol{T}_l\hat{\boldsymbol{s}}[i] =
\big{[}\hat{s}_{1,l}[i],\hat{s}_{2,l}[i],\ldots,\hat{s}_{K,l}\big{]}^T$,
a permutation of the detected symbol set $\hat{\boldsymbol{s}}[i]$,
ordered by the transformation matrix $\boldsymbol{T}_l, l =
1,\ldots,L.$, where each row and each column of $\boldsymbol{T}_l$
contain only one '1'. We also define $u_{k,l}[i]$ as the output of
the $k$-th concatenated filter for the $l$-th branch which exploits
the permutation matrix $\boldsymbol{T}_l$. The detected symbols can
be obtained in the original order by using
$\hat{\boldsymbol{s}}_l[i] =
{\boldsymbol{T}}_l^T\hat{\boldsymbol{s}}_l'[i]$. The optimal
ordering scheme conducts an exhaustive search of $L = K!$.
Sub-optimal schemes have been proposed in \cite{Rui} to design the
codebook with a reduced $L$.

\section{Iterative Detection and Decoding}
\label{sec:idd}

In the following, a soft-output detector is described to improve the
performance of the proposed detector in the concatenation with a
convolutional code. Let $b_{k,j}$ be the $j$-th bit of the
constellation symbol and ($j = 1, 2,\ldots,\log_2C$). We denote
$L[b_{k,j}]$ as the log-likelihood ratio (LLR) value for the coded
bits $b_{k,j}$. The \textit{extrinsic} information is obtained by the
detector as \cite{lsd}
\begin{equation}
L[b_{k,j}^{(e1)}] = \ln
\frac{\sum_{\boldsymbol{s}\in{\mathcal{A}_{k,j}^{1}\bigcap\mathcal{B}}}\mbox{P}\big{(}\boldsymbol{r}\big{|}\boldsymbol{s}\big{)}\exp\big{(}f(\boldsymbol{s})\big{)}}{\sum_{\boldsymbol{s}\in{\mathcal{A}_{k,j}^{0}\bigcap\mathcal{B}}}\mbox{P}\big{(}\boldsymbol{r}\big{|}\boldsymbol{s}\big{)}\exp\big{(}f(\boldsymbol{s})\big{)}}.
\end{equation}
and $\mathcal{A}_{k,j}^1$ is the set of all symbol vectors that
consist of bits satisfying $b_{k,j} = 1$, $\mathcal{A}_{k,j}^{0}$ is
similarly defined but satisfying $b_{k,j} = 0$. Similar to list-SD
\cite{lsd}, a list of vectors can be found by deploying the proposed
detector, the ML vector can be found as a tentative decision. By
appropriately selecting the tentative decisions, the AMUDFCC
detector performance can approach the optimal MLD performance. Let
$\mathcal{B}$ denote the set of tentative decisions obtained from
\begin{equation}
\mathcal{B} = \boldsymbol{B}_{1} \cup\boldsymbol{B}_{2}\cup,\ldots,\cup \boldsymbol{B}_{k}\cup,\ldots,\cup\boldsymbol{B}_{K},
\end{equation}
If $L>1$, MB is used and we have
\begin{equation}
{\mathcal{B}} = {\mathcal{B}}_{1} \cup{\mathcal{B}}_{2}\cup,\ldots,\cup {\mathcal{B}}_{l}\cup,\ldots,\cup{\mathcal{B}}_{L}.
\end{equation}
When the intersection set is empty, i.e.
${\mathcal{A}_{k,j}^{1}\cap\mathcal{B}} = \emptyset$ or
${\mathcal{A}_{k,j}^{0}\cap\mathcal{B}} = \emptyset$ the LLR for
that specific bit can be filled with an arbitrary number with a
large magnitude. The probability density can be obtained by
$\mbox{P}\big{(}\boldsymbol{r}\big{|}\boldsymbol{s}\big{)} \propto
\exp\Big{(}{-\frac{1}{\sigma_v^2}\|\boldsymbol{r}-\boldsymbol{Hs}\|^2}\Big{)},$
where $ f(\boldsymbol{s}) =
\frac{1}{2}(2\boldsymbol{b}_{[k,j]}^T-1)\boldsymbol{L}[b_{k,j}^{(p1)}]$,
where $b_{[k,j]}$ is the vector of all bits without the $j$-th bit
from the $k$-th symbol, and similarly for the L-vector.

\section{Simulation Results}
\label{sec:sim}

In this section, simulations are presented to demonstrate the system
performance of the proposed AMUDFCC detection algorithm. We consider
time-varying fading channels and QPSK modulation. The transmitted
vectors $\boldsymbol{s}[i]$ are grouped into frames of $500$ symbol
vectors where the first $10$ symbol vectors are training data and
the column-norm based ordering described in Section \ref{sec:data}
is employed.

\begin{figure}[!htb]
\begin{center}
\def\epsfsize#1#2{0.875\columnwidth}
\epsfbox{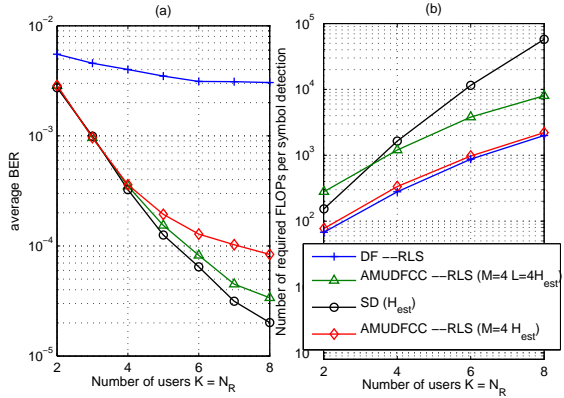} \vspace{-1.25em} \caption{\footnotesize
Performance with $E_b/N_0=13$ dB, AMUDFCC with $d_{\scriptsize
\mbox{th}} = 0.5$ and LS channel estimation. (a) AMUDFCC has a
superior performance to the conventional DF scheme and is not far
from the MLD performance obtained with the SD. (b) The AMUDFCC has a
similar cost to the conventional DF.
 }\label{twin}
\vspace{-1em}
\end{center}
\end{figure}

In Fig.\ref {twin}(a), it is shown the BER performance against the
number of users assuming $N_R=KN_T$ for a block fading channel. The
BER performances of all schemes improve while the number of receive
antenna $N_R$ grows with the number of users $K$. More importantly,
the proposed detector offers a significant performance gain over the
DF-RLS detector at a small extra computational cost as shown in
Fig.\ref {twin}(b). By adding more complexity, the performance can
be further improved by introducing $L$ parallel branches. The
computational complexity is shown in terms of floating-point
operations (FLOPS) per symbol detection. \footnote{The FLOPS were
counted by the Lightspeed toolbox \cite{lightspeed}. The FLOPS count
as 2 for a complex addition and as 6 for a complex multiplication.}

\begin{figure}[!htb]
\begin{center}
\def\epsfsize#1#2{0.875\columnwidth}
\epsfbox{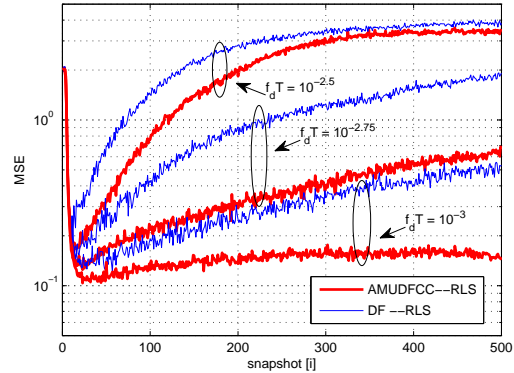} \vspace{-1.25em} \caption{\footnotesize MSE of
the estimated symbols in terms of RLS iterations, with 4 users.
After 10 training vectors, the decision-directed mode is switched
on.} \vspace{-1em} \label{MSE16}
\end{center}
\end{figure}

\begin{figure}[!htb]
\begin{center}
\def\epsfsize#1#2{0.875\columnwidth}
\epsfbox{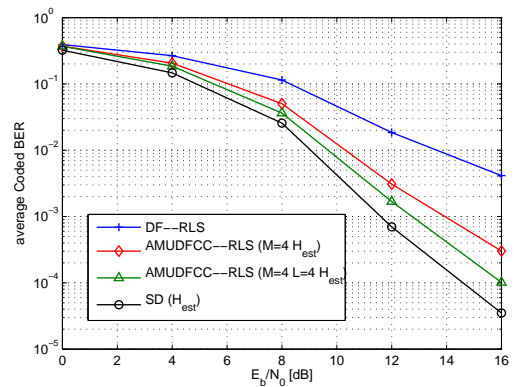} \vspace{-1.25em} \caption{\footnotesize $K=6$
users are separately coded by the $g = (7,5)_{o}$, rate $R = 1/2$,
memory 2 convolutional code and we use the block size equals 500
vectors, $M = 4$ candidates and $d_{\scriptsize \mbox{th}} = 0.5$.
The number of turbo iterations between the detector and the decoder
is 3.}\label{CodedBER} \vspace{-1.25em}
\end{center}
\end{figure}

Fig. \ref {MSE16} illustrates the MSE for the symbol estimation
across all $4$ users in terms of RLS iterations. The channel between
a transmit and receive antenna pair follows Jakes' model \cite
{lightspeed}. Here, we have $E_b/N_0=14$ dB and the normalized
Doppler frequency shift equals $10^{-2.5}$ , $10^{-2.75}$ and
$10^{-3}$, respectively. It is clear that the AMUDFCC-RLS
considerably reduces the MSE level when compared to DF-RLS. For a
coded system with RLS channel estimation, the BER performance
against the average SNR across all users is shown in
Fig.{\ref{CodedBER}}. The curves show that the proposed AMUDFCC
detector has a substantial performance gain as compared to the
conventional DF scheme. By increasing the number of branches with
different NCO, the SD performance can be approached.

\section{Conclusions}
\label{sec:con}

In this paper, we have developed an adaptive iterative decision
feedback based detector for MU-MIMO systems in time-varying channel.
The proposed scheme is able to approach the optimal MLD performance
while requiring a significantly lower computational cost.


\end{document}